\documentclass[pra,amssymb,showpacs]{revtex4}
\usepackage{graphicx,amsbsy,amsmath,bbm}
\usepackage{psfig}
\usepackage{multirow}
\usepackage{pstricks}

\input{epsf}

\newcommand{\ket}[1]{\vert #1 \rangle}
\newcommand{\bra}[1]{\langle #1 \vert}

\newcommand{\mbfrac}{\mbox{$\frac12$}}

\newcommand{\pexp}[1]{\exp\left\{ #1 \right\}}

\newcommand{\bmsigma}{\boldsymbol\sigma}

\newcommand{\bmLambda}{\boldsymbol \Lambda}

\newcommand{\bmA}{{\boldsymbol A}}
\newcommand{\bmB}{{\boldsymbol B}}
\newcommand{\bmC}{{\boldsymbol C}}

\newcommand{\bmJ}{{\boldsymbol J}}
\newcommand{\bmM}{{\boldsymbol M}}

\newcommand{\bmR}{{\boldsymbol R}}

\newcommand{\bmX}{{\boldsymbol X}}
\newcommand{\bmY}{{\boldsymbol Y}}

\newcommand{\bmZ}{{\boldsymbol Z}}

\newcommand{\calpha}{\alpha^{*}}

\newcommand{\cxi}{\xi^{*}}

\newcommand{\rmSU}{{\rm SU}}

\newcommand{\be}{\begin{equation}}
\newcommand{\ee}{\end{equation}}
\newcommand{\bea}{\begin{eqnarray}}
\newcommand{\eea}{\end{eqnarray}}

\newcommand{\sT}{\scriptscriptstyle T}

\newcommand{\ii}{\openone}
\newcommand{\iid}{\mathbb{I}}
\newcommand{\jj}{\mathbb{J}}

\newcommand{\gr}[1]{\boldsymbol{#1}}

\newcommand{\refeq}[1]{Eq.~(\ref{#1})}

\renewcommand{\det}{{\rm Det}\,}

\def\re#1{\Re\hbox{e}[#1]}
\def\im#1{\Im\hbox{m}[#1]}

\newcommand{\KPsi}{|\gr{\Psi}_m\rangle}
\newcommand{\KPsiT}{|\gr{\Psi}_2\rangle}
\newcommand{\KPhi}{|\gr{\Phi}_m\rangle}
\newcommand{\calC}{{\cal C}}
\newcommand{\bmcalN}{\boldsymbol {\cal N}}
\newcommand{\bmcalA}{\boldsymbol {\cal A}}
\newcommand{\bmcalB}{\boldsymbol {\cal B}}
\newcommand{\pp}{\mathbb{P}}
\newcommand{\GG}{\mathbb{G}}
\newcommand{\tinyT}{{\mbox{\tiny $T$}}}
\newcommand{\funf}{f(N,\gamma_0;x,\gamma_\tinyT)}

\newcommand{\ba}{\begin{align}}

\begin{document}
\title{Multimode entanglement and telecloning in a noisy environment}
\author{Alessandro Ferraro and Matteo G. A. Paris}
\affiliation{Dipartimento di Fisica dell'Universit\`a di Milano, Italy.}
\begin{abstract}
  We address generation, propagation and application of multipartite
  continuous variable entanglement in a noisy environment. In
  particular, we focus our attention on the multimode entangled states
  achievable by second order nonlinear crystals, {\em i.e.} coherent
  states of ${\rm SU}(m,1)$ group, which provide a generalization of
  the twin-beam state of a bipartite system. The full inseparability
  in the ideal case is shown, whereas thresholds for separability are
  given for the tripartite case in the presence of noise. We find that
  entanglement of tripartite states is robust against thermal noise,
  both in the generation process and during propagation. We then
  consider coherent states of $\rmSU(m,1)$ as a resource for multipartite
  distribution of quantum information, and analyze a specific protocol
  for telecloning, proving its optimality in the case of symmetric
  cloning of pure Gaussian states. We show that the proposed protocol
  also provides the first example of a completely asymmetric
  $1\rightarrow m$ telecloning, and derive explicitly the optimal
  relation among the different fidelities of the $m$ clones.  The
  effect of noise in the various stages of the protocol is taken into
  account, and the fidelities of the clones are analytically obtained
  as a function of the noise parameters. In turn, this permits the
  optimization of the telecloning protocol, including its adaptive
  modifications to the noisy environment. In the optimized scheme the
  clones' fidelity remains maximal even in the presence of losses (in
  the absence of thermal noise), for propagation times that diverge as
  the number of modes increases. In the optimization procedure the
  prominent rule played by the location of the entanglement source
  is analyzed in details. Our results indicate that, when only losses are
  present, telecloning is a more effective way to distribute quantum
  information then direct transmission followed by local
  cloning.
\end{abstract}
\maketitle
\section{Introduction}\label{intro}
Entanglement plays a fundamental role in quantum information, being
recognized as the essential resource for quantum computing,
teleportation, and cryptographic protocols. In the framework of
quantum information with continuous variables (CV)
\cite{vLB_rev,Napoli} the possibility of generating and manipulating
entanglement allowed the realization of a variety of quantum
protocols, such as teleportation, cryptography, dense coding and
entanglement swapping. In these protocols the source of entanglement
is the bipartite twin-beam state of two modes of radiation, usually
generated by parametric down-conversion in $\chi^{(2)}$ crystals.
However, recent experimental progresses \cite{Exps} show that the
coherent manipulation of entanglement between more then two modes may
be achieved with current technology. This opens the opportunity to
realize a true quantum information network, in which the information
can be stored, manipulated and distributed among many parties, in a
fashion resembling the current classical telecommunication networks.
In a realistic implementation, entanglement needs to be transmitted
along physical channels, such as optical fibers or the atmosphere.  As
a matter of fact, the propagation and the influence of the environment
unavoidably lead to degradation of entanglement, owing to decoherence
effects induced by losses and thermal noise. In this scenario, it is
worth to study the entanglement properties and the possible
applications of multipartite systems in noisy environments, which will
be the subject of this paper.
\par
A prominent class of CV states is constituted by the Gaussian states.
They can be theoretically characterized in a convenient way, and
generated and manipulated experimentally in a variety of physical
systems. In a quantum information setting, entangled Gaussian states
provide the basis for the quantum information protocols mentioned
above. The basic reason for this is that the QED vacuum and radiation
states at thermal equilibrium are themselves Gaussian states. This
observation, in combination with the fact that the evolutions
achievable with current technology are described by Hamiltonian
operators at most bilinear in the fields, accounts for the fact that
the states commonly produced in labs are Gaussian. Indeed, bilinear
evolutions preserve the Gaussian character. As we already mentioned,
the outmost used source of CV entanglement are the twin-beams, which
belong to the class of bipartite Gaussian states. In a
group-algebraic language, they are the coherent states of the group
$\rmSU(1,1)$, {\em i.e.}, the states evolved from vacuum via a unitary
realization of the group. Within the class of Gaussian states, the
simplest generalization of twin-beams to more then two modes are the
coherent states of the group $\rmSU(m,1)$. Indeed, this states can be
generated by multimode parametric processes in second order nonlinear
crystals, with Hamiltonians that are at most bilinear in the fields.
In particular, these processes involve $m+1$ modes of the field
$a_0,a_1,\dots,a_m$, with mode $a_0$ that interacts through a
parametric-amplifier-like Hamiltonian with the other modes, whereas
the latter interact one each other only via a beam-splitter-like
Hamiltonian \cite{vecchio,puri}. In the framework of CV quantum
information, the first proposal to produce such states has been given
in Ref.~\cite{vLB_tlc}, where a half of a two-mode squeezed vacuum
state interacts with $m$ vacua via a proper sequence of beam
splitters. Other unitary realizations of the algebra of $\rmSU(m,1)$
have been proposed, in optical settings \cite{como,chirkin} as well as
with cold atoms \cite{nics} or optomechanical systems \cite{cams}. In
these schemes the Hamiltonian of the system, rather then involving a
sequence of two-mode interaction, is realized via simultaneous
multimode interactions. Experimental realizations of tripartite
entanglement in the optical domain have been recently reported
\cite{Exps}.
\par
In this work we do not focus on any specific implementation of the
$\rmSU(m,1)$ evolution. Rather, we will analyze the entanglement
properties of $\rmSU(m,1)$ coherent states in a unified fashion valid
for a generic Hamiltonian of this kind. As we will see in
Sec.~\ref{s:SUm1}, this is allowed by the observation that the
coherent states of $\rmSU(m,1)$ have a common structure, which can be
conveniently written in the Fock representation of the field
\cite{puri}. In particular, the degradation effects of both the
thermal background in the generation process and of losses and thermal
photons in the propagation will be outlined. The
robustness of these states against noise will be analyzed in
Sec.~\ref{s:NoisySUm1} where it will be also compared with the
bipartite case.
\par
As already mentioned, one of the main results in CV quantum
communication concerned the realization of the teleportation protocol
(for a recent experiment see \cite{furu05}). The natural
generalization of standard teleportation to many parties corresponds
to a telecloning protocol \cite{murao}.  Teleportation is based on the
coherent states of $\rmSU(1,1)$, which provide the shared entangled
states supporting the protocol. Thus, in order to implement a
multipartite version of this protocol, one is naturally led to
consider a shared entangled state produced by a generic $\rmSU(m,1)$
interaction. The telecloning protocol will be analyzed in detail in
Sec.~\ref{s:tlc}. As concern cloning with CV, there are general
results to assess the optimality of $n\rightarrow m$ symmetric cloning
of coherent states \cite{cerf}.  Optimal local unitary realization of
such schemes have been proposed in \cite{braunetal,fiurasek}, and an
experimental realization of $1\rightarrow 2$ cloning has been recently
reported \cite{leuchs}.  Concerning telecloning, existing proposals
are about optimal $1\rightarrow m$ symmetric cloning of pure Gaussian
states, using a particular coherent state of $\rmSU(m,1)$ as support
\cite{vLB_tlc}.  Recently, a proposal which make use of partially
disembodied transport has also been reported \cite{zhang}. In view of
the realization of a quantum information network, one is naturally led
to consider the possibility to retrieve different amount of
information from different clones.  This means that one may consider
the possibility to produce clones different one from each other, in
what is called asymmetric cloning.  Examples of optimal $1\rightarrow
2$ asymmetric cloning are given in Ref.~\cite{fiurasek,josab}, where
local and non-local realizations are considered. In this work, we will
see how the telecloning protocol involving a generic coherent state of
$\rmSU(m,1)$ provides the first example of a completely asymmetric
$1\rightarrow m$ cloning of pure Gaussian states. In this sense, we
provide a generalization of the proposal in Ref.~\cite{vLB_tlc} to the
asymmetric case. Moreover, we found an expression for the maximum
fidelity achievable by one clone when the fidelities of the others are
fixed to prescribed values, thus giving explicitly the trade-off
between the qualities of the different clones.
\par
In Sec.~\ref{s:NoisyTLC} we will analyze the effect of noise in each
step of the telecloning protocol. As expected, the presence of both
thermal noise and losses unavoidably leads to a degradation of the
cloning performances. Nevertheless, we will show that the protocol can
be optimized in order to reduce these degradation effects. In
particular, one may optimize not only the energy of the entangled
support, but also the location of the source of entanglement itself.
Remarkably, when only losses are considered, this optimization
completely cancels the degrading effects of noise on the fidelity of
the clones. This happens for finite propagation times which, however,
diverge as the number of modes increases.
\par
We conclude the paper with Sec.~\ref{esco}, where the main results
will be summarized.
\section{Multimode parametric
  interactions: ${\rm\bf SU} \boldsymbol{(m,1)}$
  coherent states}\label{s:SUm1}
Let us consider the set of bilinear Hamiltonians expressed by
\begin{equation}
H_m=\sum_{l<k=1}^{m} \gamma_{kl}^{(1)}\, a_k a^{\dag}_l
+ \sum_{k=1}^{m} \gamma_{k}^{(2)}\, a_k a_0 + h. c.
\label{Hm}\;,
\end{equation}
where $[a_k,a_l]=0$, $[a_k,a^\dagger_l]=\delta_{k,l}$
($k,l=0,\dots,m$) are independent bosonic modes. A conserved quantity
is the difference $D$ between the total mean photon number of the mode
$a_0$ and the remaining modes, in formula
\begin{eqnarray}
D= \sum_{k=1}^m a^\dag_k a_k - a_0^\dag a_0
\label{cons}\;.
\end{eqnarray}
The transformations induced by Hamiltonians (\ref{Hm}) correspond to
the unitary representation of the $\rmSU(m,1)$ algebra \cite{puri}.
Therefore, the set of states obtained from the vacuum coincides with
the set of $\rmSU(m,1)$ coherent states {\em i.e.}
\begin{eqnarray}
\KPsi\equiv
\pexp{-i H_{m} t} |\gr{0}\rangle = \pexp{\sum_{k=1}^m
\beta_{k} a_k^\dag a_0^\dag - h.c.} |\gr{0}\rangle
\label{PsiAux1}\;,
\end{eqnarray}
where $\beta_{k}$ are complex numbers, parameterizing the state, which
are related to the coupling constants $\gamma_{kl}^{(1)}$ and
$\gamma_{k}^{(2)}$ in \refeq{Hm}. Upon defining
$$\calC_{k} = \beta_{k}\frac{\tanh\left(\sum_{r=1}^m
|\beta_{r}|^2\right)}{\sum_{r=1}^m |\beta_{r}|^2}\:,$$
$\KPsi$ in Eq.~(\ref{PsiAux1}) can be explicitly
written as
\begin{align}
\KPsi =& \sqrt{{\cal Z}_m}\sum_{\{\gr{n}\}}
\frac{\calC_1^{n_1} \calC_2^{n_2}... \calC_m^{n_m}\: \sqrt{(n_1+n_2+...
+n_m)!}}{\sqrt{n_1! n_2! ... n_m!}}\:
|\sum_{k=1}^m n_k ;\{\gr{n}\} \rangle
\label{Psi}\;,
\end{align}
where $\{\gr{n}\}=\{n_1,n_2,...,n_m\}$. The sums over $\gr{n}$ are
extended over natural numbers and ${\cal Z}_m= 1-\sum_{k=1}^m
|\calC_k|^2$ is a normalization factor. We see that for $m=1$ one
recover the twin-beam state. Notice that, being interested in the
entanglement properties and applications of states $\KPsi$, we can
take the $\calC_k$'s coefficients as real numbers. In fact one can put
to zero the possible phases associated to each $\calC_k$ by performing
a proper local unitary operation on mode $a_k$, which in turn does not
affect the entanglement of the state.  Calculating the expectation
values of the number operators $N_k=\langle a^\dag_k a_k \rangle$ on
the multipartite state $\KPsi$ one may re-express the coefficient in
\refeq{Psi} as follows:
\begin{align}
\calC_k=\left(\frac{N_k}{1+N_0}\right)^{1/2} \;,
\qquad {\cal Z}_m=\frac{1}{1+N_0}
\qquad (k=1,\dots m)
\label{calCs}\;.
\end{align}
In order to obtain \refeq{calCs} we have considered \refeq{cons} with
$D=0$ (vacuum input), from which follows that 
\begin{align}
  N_0=\sum_{k=1}^m N_k \;,
\label{cons2}
\end{align}
and repeatedly used the following identity: \ba \sum_{n=0}^\infty
x^n\frac{(n+a)!}{n!}=a!(1-x)^{-1-a} \;.
\label{SeriesIdentity}
\end{align}
The case $D\neq0$ will be considered in the next section, in which the
effects of thermal background will be taken into account. The basic
property of states in \refeq{Psi} is their full inseparability, {\em
  i.e.}, they are inseparable for any grouping of the modes. To prove
this statement first notice that, being evolved with a bilinear
Hamiltonian from the vacuum, the states $\KPsi$ are pure Gaussian
states. They are completely characterized by their covariance matrix
$\bmsigma$, whose entries are defined as
\begin{align}
\label{defCOV}
[\bmsigma]_{kl} &=  \frac12 \langle \{R_k,R_l\} \rangle -
\langle  R_l \rangle
\langle  R_k \rangle\,,
\end{align}
where $\{A,B\}=AB+BA$ denotes the anticommutator, $\bmR=
(q_0,p_0,\ldots,q_m,p_m)^{\sT}$ and the position and momentum operator
are defined as $q_k = (a_k + a_k^\dag)/\sqrt2$ and $p_k = (a_k -
a_k^\dag)/\sqrt2$.  The covariance matrix for the states $\KPsi$ reads
as follows:
\begin{align}
\label{CovPsi}
\bmsigma_{m} &= \left(
\begin{array}{ccccc}
\bmcalN_0 & \bmcalA_1     & \bmcalA_2     & \ldots          &  \bmcalA_m \\
\bmcalA_1 & \bmcalN_1     & \bmcalB_{1,2} & \ldots          &  \bmcalB_{1,m} \\
\bmcalA_2 & \bmcalB_{1,2} & \bmcalN_2     & \ddots          &  \vdots \\
\vdots    & \vdots        & \ddots        & \ddots          &  \bmcalB_{m-1,m} \\
\bmcalA_m & \bmcalB_{1,m} & \ldots        & \bmcalB_{m-1,m} &  \bmcalN_m \\
\end{array}
\right)\,,
\end{align}
where the entries are given by the following $2\times 2$ matrices
($k=0,\dots,m$, $h=1,\dots,m$, $j=2,\dots,m$ and $0 < i < j$)
\begin{align}
\label{CovPsiAux}
\bmcalN_k=(N_k+\frac12)\,\ii \qquad
\bmcalA_h= \sqrt{N_h(N_0+1)}\,\pp \qquad
\bmcalB_{i,j} = \sqrt{N_i\,N_j}\,\ii \;,
\end{align}
with $\ii={\rm Diag}(1,1)$ and $\pp={\rm Diag}(1,-1)$.
Since $\KPsi$ are pure states, full inseparability can be demonstrated
by showing that the Wigner function does not factorize for any
grouping of the modes, which in turn is ensured by the explicit
expression of the covariance matrix $\bmsigma_m$ given above (as soon
as $N_h\neq0$).
\section{Effect of noise on the generation and propagation of ${\rm\bf SU} \boldsymbol{(m,1)}$
  coherent states} \label{s:NoisySUm1} 
In view of possible applications of the coherent states of $\rmSU
(m,1)$ to a real quantum communication scenario, it is worth to
analyze the degrading effects on their entanglement that may arise
when generation and propagation in a noisy environment is taken into
account.  Unfortunately, a manageable necessary and sufficient
entanglement criterion for the general case of a Gaussian multipartite
state is still lacking. Thus, in order to study quantitatively the
effects of noise we must limit ourselves to the case when only three
modes are involved (insights for the general $m$-mode case will be
given in the following Sections). In fact, up to three modes the
partial transpose criterion introduced in \cite{ppt,duan,simon} is
necessary and sufficient for separability \cite{ppt3}. It says that a
Gaussian state described by a covariance matrix $\bmsigma$ is fully
inseparable if and only if the matrices
$\omega_k=\bmsigma-\frac{i}{2}{\widetilde \bmJ}_k$ ($k=0,1,2$) are
non-positive definite, where ${\widetilde \bmJ}_k=\bmLambda_k \bmJ
\bmLambda_k$ with $\bmLambda_0={\rm Diag}(1,-1,1,1,1,1)$, $\bmLambda_1
={\rm Diag}(1,1,1,-1,1,1)$, $\bmLambda_2 ={\rm Diag}(1,1,1,1,1,-1)$
and
\begin{align}
\label{InsepMat}
 \gr{J} = \left(\begin{array}{cc} \boldsymbol{0} &-
\ii_3 \\
\ii_3
&\boldsymbol{0} \end{array}\right)\;,
\end{align}
and $\ii_n$ is the $n\times n$ identity matrix. This criterion has
been applied in Refs.~\cite{ppt3,chen} in order to assess the
separability of the CV tripartite state proposed in
Ref.~\cite{vLB_tlpnet} when thermal noise is taken into account. In
Ref.~\cite{cams} the entanglement properties of a state generated via
a $\rmSU (2,1)$ evolution when one of the modes starts from thermal
background has been also numerically addressed.
\par
Let us now analyze if the generation process of states $\KPsi$ is
robust against thermal noise. This means that we have to study the
separability properties of a state generated by a $\rmSU (m,1)$
interaction starting from a thermal background rather then from the
vacuum, in formulae $\varrho=e^{-iH_mt}\varrho_\nu\, e^{iH_mt}$, where
$\varrho$ and $\varrho_\nu$ are the density matrix of the evolved
state and of a thermal state, respectively. We may call these states
thermal coherent states of $\rmSU (m,1)$. First notice that, being the
thermal state Gaussian, the thermal coherent states will be Gaussian
too, and their covariance matrix $\bmsigma_{m,{\rm th}}$ may be
immediately identified from \refeq{CovPsi}. In fact, in the phase
space identified by the vector $\bmR$, every $\rmSU (m,1)$ evolution
will act as a symplectic operation ${\boldsymbol {\cal S}}$ on the
covariance matrix of the input state, {\em i.e.}, $\bmsigma_{\rm
  out}={\boldsymbol {\cal S}}^T\,\bmsigma_{\rm in}{\boldsymbol {\cal
    S}}$. Recalling that the covariance matrix of a thermal state can
be written as $\bmsigma^{\rm th}=(2\,\nu+1)\bmsigma_v$, being
$\bmsigma_v=\ii/2$ the covariance matrix of vacuum and $\nu$ the mean
thermal photon number, we obtain
\begin{align}
\label{sigma_th}
  \bmsigma_{m,{\rm th}}=(2\nu+1)\bmsigma_m
\end{align}
Let us now apply the separability criterion recalled above to
$\bmsigma_{3,{\rm th}}$. Concerning the first mode,
from an explicit calculation of the minimum eigenvalue of matrix
$\omega_0$ it follows that
\begin{equation}
\lambda^{\rm min}_0= \nu+(1+2 \nu)
\left[N_0-\sqrt{N_0(N_0+1)}\right]\ \;.
\label{c4:TThAvalMin}
\end{equation}
As a consequence, mode $a_0$ is separable from the others when
\begin{equation}
\nu > N_0+\sqrt{N_0(N_0+1)} \;.
\label{c4:TThTrsh1}
\end{equation}
Calculating the characteristic polynomial of matrix $\omega_1$ one
deals with the following pair of cubic polynomials
\begin{multline}
q_1(\lambda,N_0,N_1,N_2,\nu) =
\lambda^3 - 2\left[ 2(1+ N_0)+ \nu(3+4N_0)
\right]\lambda^2\\
+4\left[1+ N_1+2N_2+
 \nu(4+4N_1+6N_2+ \nu (3+4N_0))\right]\lambda\\
-8 \nu\left[1+N_1+ \nu(2+ \nu
+2N_1)\right]
\;,
\end{multline}
\vspace{-1cm}
\begin{multline}
q_2(\lambda,N_0,N_1,\nu) =
\lambda^3
-2\left[1+2N_0+ \nu(3+4N_0) \right]\lambda^2
\\
+4\left[ N_1+2 \nu(1+N_0)+ \nu^2(3+4N_0)
\right]\lambda\\
-8(1+ \nu)( \nu^2-2N_1-2\nu N_1)
\;.
\end{multline}
While the first polynomials admits only positive roots, the second one
shows a negative root under a certain threshold. It is possible to
summarize the three separability thresholds of the three modes
involved in the following inequalities
\begin{equation}
\nu > N_k+\sqrt{N_k(N_k+1)} \;.
\label{c4:TThGenericThrs}
\end{equation}
If Inequality (\ref{c4:TThGenericThrs}) is satisfied for a given $k$,
then mode $a_k$ is separable. Clearly, it follows that the state
$\KPsiT$ evolved from vacuum ({\em i.e.}, $ \nu=0$) is fully
inseparable, as expected from Section \ref{s:SUm1}. Remarkably,
Inequality (\ref{c4:TThGenericThrs}) is the same as for the twin beam
evolved from noise \cite{serale}, which means that the entanglement of
the thermal coherent states of $\rmSU(2,1)$ is as robust against noise
as it is for the case of the thermal coherent states of $\rmSU(1,1)$.
\par
Let us now consider the evolution of the state $\KPsiT$ in three
independent noisy channels characterized by loss rate $\Gamma$ and
thermal photons $\mu$, equal for the three channels. The covariance
matrix $\bmsigma_2(t)$ at time $t$ is given by a convex combination of
the ideal $\bmsigma_2(0)$ [{\em i.e.}, $\bmsigma_2$ in \refeq{CovPsi}]
and of the stationary covariance matrix
$\bmsigma_{\infty,2}=(\mu+\mbfrac)\ii_6$
\begin{equation}
\bmsigma_2(t)=e^{-\Gamma t}\,\bmsigma_2+ (1-e^{-\Gamma t})\,\bmsigma_{\infty,2}
\;.
\label{c4:3mCMEvol}
\end{equation}
Consider for the moment a pure dissipative environment, namely
$\mu=0$.  Applying the separability criterion above to
$\bmsigma_2(t)$, one can show that it describes a fully inseparable
state for every time $t$. In fact, we have that the minimum eigenvalue
of $\omega_0$ is given by
\begin{equation}
\lambda^{\rm min}_0= 2e^{-\Gamma t}\left[N_0-\sqrt{N_0(N_0+1)}\right]\,.
\label{c4:LChAvalMin1}
\end{equation}
Clearly, $\lambda_0^{\rm min}$ is negative at every time $t$, implying
that mode $a_0$ is always inseparable from the others. Concerning mode
$a_1$, the characteristic polynomial of $\omega_1(t)$ factorizes into two
cubic polynomials:
\begin{subequations}
\label{c4:LossyChCubic}
\begin{align}
q_1(\lambda,\Gamma,N_0,N_1,N_2) &=
-\lambda^3+4\left[ 1+ e^{-\Gamma t}N_0 \right]\lambda^2 \nonumber\\
&\hspace{1cm}+4\left[ -1-e^{-\Gamma t}( 2N_1 + 3N_2
- e^{-\Gamma t} N_0) \right]\lambda + 8e^{-\Gamma t}N_2(1-e^{-\Gamma t})\;,
\\
q_2(\lambda,\Gamma,N_0,N_1,N_2) &=
-\lambda^3 + 2\left[1+2e^{-\Gamma t}N_0 \right]\lambda^2 \nonumber\\
&\hspace{3cm}
+ 4\left[ -e^{-\Gamma t}(2N_1+N_2)+ e^{-2\Gamma t}N_0 \right]\lambda
- 8e^{-2\Gamma t}N_1 \;.
\end{align}
\end{subequations}
While the first polynomial has only positive roots, the second one
admits a negative root at every time. Due to the symmetry of state
$\KPsiT$ the same observation applies to mode $a_2$, hence full
inseparability follows. This result resembles again the case of the
twin beam state in a two-mode channel \cite{duan,kim_stefano}. In other
words, the behavior of the coherent states of $\rmSU(2,1)$ in a pure
lossy environment is the same as the behavior of the coherent states
of $\rmSU(1,1)$, concerning their entanglement properties.
\par
\begin{figure}[b]
\includegraphics[width=5cm]{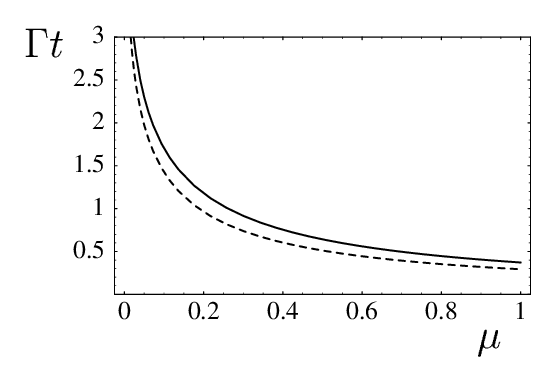}
\caption{Separability thresholds for modes $a_0$ (continuous line) and $a_1$ (dashed line) according to \refeq{threshold0} and \refeq{threshold1} for the case of $N_1=N_2=N=1$. The behavior of these curves is similar if different values of $N$ are considered.} \label{f:thresh}
\end{figure}
When thermal noise is taken into account ($\mu\neq0$) separability
thresholds arise, which again resembles the two-mode channel case. Concerning
mode $a_0$, the minimum eigenvalue of matrix $\omega_0(t)$ is
negative when
\be
\label{threshold0}
t < \frac{1}{\Gamma}
\ln\left(1+\frac{\sqrt{\frac{1}{2}N_{tot}(\frac{1}{2}N_{tot}+1)}-\frac{1}{2}N_{tot}}{\mu}\right)
\;, \ee where $N_{tot}=N_0+N_1+N_2$.  
Remarkably, this threshold is exactly the same as the two-mode case \cite{duan}, if
one consider both of them as a function of the total mean photon
number of the TWB and of state $\KPsiT$ respectively.  This
consideration confirms the robustness of the entanglement of the
tripartite state $\KPsiT$. Concerning mode $a_1$, the characteristic
polynomial of $\omega_1(t)$ factorizes again into two
cubic polynomials. As above, one of the two have always positive
roots, while the other one admits a negative root for time $t$ below a
certain threshold, in formulae:
\begin{multline}
-8e^{-2\Gamma t}N_1+8(e^{-\Gamma t}-1)e^{-\Gamma t}(e^{-\Gamma t} N_0
- 2N_1 - N_2)\mu \\
+8(e^{-\Gamma t}-1)^2(1+2e^{-\Gamma t}N_0)\mu^2
- 8(e^{-\Gamma t}-1)^3 \mu^3<0 \;.
\label{threshold1}
\end{multline}
Mode $a_2$ is subjected to an identical separability threshold, upon
the replacement $N_1\leftrightarrow N_2$. In Fig.~\ref{f:thresh} we
compare the separability thresholds given by \refeq{threshold0} and
\refeq{threshold1}. As it is apparent from the plot, modes $a_1$ and
$a_2$ become separable faster then modes $a_0$, hence the threshold
for full inseparability of $\KPsiT$ is given by \refeq{threshold1}.
\par
We conclude that the entanglement properties of the coherent states of
$SU(2,1)$ in a noisy environment resembles the twin-beam case both in
generation and during propagation. This may be relevant for
applications, as the robustness of twin beam is at the basis of
current applications in bipartite CV quantum information.
\section{Telecloning}\label{s:tlc}
\begin{figure}[t]
\includegraphics[width=10.9cm]{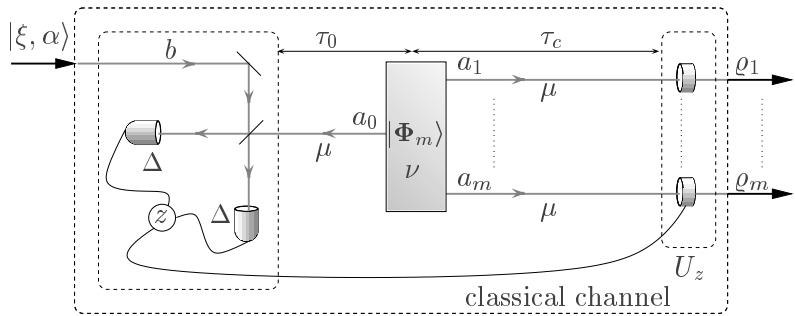}
\caption{Schematic diagram of the telecloning scheme. After the
  preparation of the state $\KPhi$, a conditional measurement is made
  on the mode $a_0$, which corresponds to the joint measurement of the
  sum- and difference-quadratures on two modes: mode $a_0$ itself and
  another reference mode $b$, which is excited in a pure Gaussian
  state $|\xi,\alpha\rangle$, to be teleported and cloned.  The result
  $z$ of the measurement is classically sent to the parties who want
  to prepare approximate clones, where suitable displacement
  operations (see text) on modes $a_1,\dots,a_m$ are performed. We
  indicated with $\nu$ and $\mu$ the mean thermal photons in
  generation and propagation, whereas $\Delta$ takes into account for
  the non unit efficiency in the detection stage. The effective
  propagation times $\tau_0$ and $\tau_c$ (see Sec.~\ref{s:NoisyTLC})
  are related to the losses during propagation.
\label{f:tlc}}
\end{figure}

We now show how the multipartite states $\KPsi$ introduced in
Sec.~\ref{s:SUm1} can be used in a quantum communication scenario.  In
particular we show that states $\KPsi$ permit to achieve optimal
symmetric and asymmetric $1\rightarrow m$ telecloning of pure Gaussian
states. Optimal symmetric telecloning has been in fact already
proposed in \cite{vLB_tlc} using a shared state produced by a
particular $\rmSU(m,1)$ interaction. Also a protocol performing
optimal $1\rightarrow 2$ asymmetric telecloning of coherent states has
been already suggested in Ref.~\cite{josab}, where the shared state is
produced by a suitable bilinear Hamiltonian which generates a
$\rmSU(2,1)$ evolution operator. Here we consider the general
$1\rightarrow m$ telecloning of Gaussian pure states in which the
shared entanglement is realized by a generic coherent state of $\rmSU
(m,1)$. First recall that a single-mode pure Gaussian state can be
always written as
\begin{align}
\label{1Mpure}
|\xi,\alpha\rangle &= S_b(\xi)\,D_b(\alpha)|0\rangle \;,
\end{align}
where $S_b(\xi)= \pexp{\frac12 \xi (b^\dag)^2-\frac12\cxi b^2}$ and
$D_b(\alpha)=\pexp{\alpha b^\dag -\calpha b}$ are the squeezing and
the displacement operator respectively, whereas $b$ is the mode to be
cloned. We emphasize that our goal is to create $m$ clones of state
$\varrho_{\rm in}=|\xi,\alpha\rangle\langle\xi,\alpha|$ in a non-universal
fashion, {\em i.e.} the information that we clone is encoded only in
the coherent amplitude $\alpha$. In other words, we consider the
knowledge of the squeezing parameter $\xi$ as a part of the protocol,
as in the case of local cloning of Gaussian pure states \cite{cerf23}.
The telecloning protocol is schematically depicted in
Fig.~\ref{f:tlc}.  As a shared entangled state we consider the
following \cite{note1}:
\begin{align}
\label{shared}
\KPhi=S_{a_0}(\cxi)\otimes S_{a_1}(\xi) \otimes \ldots \otimes S_{a_m}(\xi) \KPsi \;.
\end{align}
After the preparation of the state $|\gr{\Phi}_m\rangle$, a joint
measurement is made on modes $a_0$ and $b$, which corresponds to measure 
the complex photocurrent $Z = b + a_0^\dag$ (double-homodyne
detection), as in the teleportation protocol. The 
measurement is described by the POVM $\{\Pi(z)\}_{z\in {\mathbb C}}$, 
acting on the mode $a_0$, whose elements are given by \ba
\Pi (z) &= \pi^{-1} D_{a_0}(z)\: \varrho_{\rm in}^{\sT} D_{a_0}^\dag (z) \nonumber \\
&= \pi^{-1} S_{a_0}(\cxi)D_{a_0}(z')D_{a_0}(\calpha)\,
\ket{0}\bra{0}\, D^\dag_{a_0}(\calpha) D_{a_0}^\dag
(z')S^\dag_{a_0}(\cxi) \;,
\label{povm}
\end{align}
where $z$ is the measurement outcome, $z'=z \cosh r +e^{-i\theta}
z^{*} \sinh r $, $\xi=re^{i\theta}$ and $^T$ denotes transposition.
The probability distribution of the outcomes is given by
\begin{align}
P (z) &= \hbox{Tr}_{0\dots m} \left[\KPhi\langle\gr{\Phi}_m|\:
\Pi(z) \bigotimes_{h=1}^m
\iid_h \right] \nonumber \\
&= \frac{1}{\pi(1+N_0)} \: \exp
\left\{-\frac{|z'+\alpha^*|^2}{1+N_0}\right\}
\label{Pz}\;,
\end{align}
being $\iid_h$ the identity operator acting on mode $a_h$.
The conditional state of the remaining modes then reads
\begin{align}
\varrho_z &= \frac{1}{P(z)}\: \hbox{Tr}_{0}
\left[\KPhi\langle\gr{\Phi}_m|\:
\Pi(z) \bigotimes_{h=1}^m
\iid_h \right] \nonumber \\
&= \bigotimes_{h=1}^m S_{a_h}(\xi)|\calC_h(z'^{*}+\alpha)\rangle
\langle\calC_h(z'^{*}+\alpha)| S^\dag_{a_h}(\xi) \label{RhoZ}\;,
\end{align}
where $|\calC_h(z'^{*}+\alpha)\rangle$ denotes a coherent state (of
the usual Heisenberg Weyl group) with amplitude
$\calC_h(z'^{*}+\alpha)$. After the measurement, the conditional state
should be transformed by a further unitary operation, depending on the
outcome of the measurement. In our case, this is a $m$-mode product
displacement $U_z = \bigotimes_{h=1}^m D_h^{\sT}(z)$. This is a local
transformation, which generalizes to $m$ modes the procedure already
used in the original CV teleportation protocol. The overall state is
obtained by averaging over the possible outcomes
$$
\varrho_{1\dots m}=\int_{\mathbb C} d^2 z\: P (z) \:
\tau_z\:.$$ where $\tau_z=U_z\: \varrho_z\:
U_z^\dag$. Thus, the partial traces
$\varrho_h=\hbox{Tr}_{1,\dots,h-1,h+1,\dots,m}[\varrho_{1\dots m}]$ read as follows
\begin{align}
\varrho_h = S_h(\xi)\,\left[\int_{\mathbb C} d^2z\: P (z) \:
|\alpha\,\calC_h+z'^*\,(\calC_h-1)\rangle\langle
\alpha\,\calC_h+z'^*\,(\calC_h-1) |\,\right] S^\dag_h(\xi)\;.
\end{align}
Upon a changing in the integration variable we obtain the following
expression for the clones:
\begin{align}
\label{clonesAUX}
\varrho_h = S_h(\xi)\,\left[\int_{\mathbb C} d^2w\:\frac{1}{\pi\, n_h}\exp \left\{ -\frac{|w-\alpha|^2}{n_h} \right\}  \:
|w\rangle\langle
w |\,\right] S^\dag_h(\xi)\;,
\end{align}
where we defined
\begin{align}
\label{noise}
n_h=\left(\sqrt{N_0+1}-\sqrt{N_h}\right)^2\,.
\end{align}
From expression (\ref{clonesAUX}) one immediately recognize that
the clones are given by thermal states $\varrho_{\rm th}(n_h)$, with mean
photon number $n_h$, displaced and squeezed by the amounts $\alpha$ and
$\xi$ respectively, {\em i.e.}:
\begin{align}
\label{clones}
\varrho_h=
S_h(\xi)\,D_h(\alpha)\,\varrho_{\rm th}(n_h)\,D^\dag_h(\alpha)\,S^\dag_h(\xi)\;.
\end{align}
As a consequence, we see that the protocol acts like a proper
covariant Gaussian cloning machine \cite{cerf23}, and that the noise
introduced by the cloning process is entirely quantified by the
thermal photons $n_h$, which in turn depend only on the value of the
mean photon numbers $N_h$ of the shared state. The fidelity $F_h$
between the $h$-th clone and the initial state $|\xi,\alpha\rangle$
does not depend on the latter and is given by
\begin{align}
F_h=\frac{1}{1+n_h}\,.
\label{CloFid}
\end{align}
\par
The expression of the clones in \refeq{clones} says that they can
either be equal or different one to each other, depending on the
values of the $n_h$'s. In other words, a remarkable feature of this
scheme is that it is suitable to realize both symmetric, when
$n_2=\dots=n_m=n$, and asymmetric cloning, $n_2\neq \dots \neq n_m$.
This arises as a consequence of the possible asymmetry of the state
that supports the telecloning. To our knowledge, this is the first
example of a completely asymmetric $1\rightarrow m$ cloning machine
for continuous variable systems.
\par
Concerning the symmetric cloning one has that this scheme saturates
the bound given in Ref.~\cite{cerf}, hence ensuring the optimality of
the protocol. In fact, the minimum added noise for a symmetric
$1\rightarrow m$ cloner of coherent states is given by
$n=\frac{m-1}{m}$, which in our case can be attained by setting
$N_1=\dots=N_m=N^{\rm opt}$, where
\begin{align}
  N^{\rm opt}=\frac{1}{m(m-1)}
\label{NOptId}
\end{align}
It follows that the fidelity is optimal, namely $F=m/(2m-1)$.  It is
not surprising that this result is the same as the one obtained in
Ref.~\cite{vLB_tlc}. In fact, as already mentioned, the latter uses
as support a specific $\rmSU (m,1)$ coherent state, generated with a particular
interaction built from single mode squeezers and beam-splitters. Our
calculation extends this result to any $\rmSU (m,1)$ coherent state
used to support the telecloning protocol.
\par
\subsection{Asymmetric cloning} \label{ss:asym}
Consider now the case of asymmetric cloning. In this case one deals with a true quantum
information distributor, in which the information encoded in an
original state may be distributed asymmetrically between many parties
according to the particular task one desires to attain. In this
scenario, a particularly relevant question concerns the
maximum fidelity achievable by one party, say $F_1$, once the
fidelities $F_j$ ($j=2,\dots,m$) of the other ones are fixed. Thanks
to \refeq{CloFid} we see that this is equivalent to the issue of
finding the minimum noise $n_1$ introduced by the cloning process for
fixed $n_j$'s ($n_j\neq1$). The optimization has to be performed under the
constrain given by \refeq{cons}, which allows to write $n_1$ as a
function of the $n_j$'s and of the total mean photon number $N_0$ (the
sums run for $j=2,\dots,m$):
\begin{align}
  \label{n1}
n_1=\left[\sqrt{N_0+1}-\sqrt{N_0-\sum_j\left(\sqrt{N_0+1}-\sqrt{n_j}\right)^2}\right]^2
\;.
\end{align}
The minimum noise $n_1^{\rm min}$ is then
found setting $N_0$ such that
\begin{align}
\label{EqA}
(N_0+1)(m-1)(m-2)-2\,\sqrt{N_0+1}(m-1){\scriptstyle \sum_j}\sqrt{n_j}
+{\scriptstyle \sum_j}n_j
+\left({\scriptstyle \sum_j}\sqrt{n_j}\right)^2-1=0
\,.
\end{align}
For $m=2$ one obtains that the optimal choice for $N_0$ is given by $N_0^{\rm
  opt}=n_2+1/4n_2$. It follows that the minimum noise $n_1^{\rm min}$
allowed by our telecloning protocol for fixed $n_2$ is given by
$n_1^{\rm min}=1/4n_2$. Hence we recover the result of
Ref.~\cite{josab} for the fidelities:
\begin{align}
  \label{Fid12}
F_1^{\rm max}=\frac{4(1-F_2)}{4-3F_2}\;.
\end{align}
Notice that if one requires $F_2=1$ then $F_1=0$, that is no
information is left to prepare a non-trivial clone on mode $a_1$. We
remark that the result in \refeq{Fid12} shows that the protocol introduced above,
besides reaching the optimal bound in the symmetric case, is optimal
also in the case of asymmetric $1\rightarrow 2$ cloning
\cite{fiurasek}. Coming back to the general case we see from
Eqs.(\ref{n1}) and (\ref{EqA}) that for $m\ge 3$ the minimum noise
$n_1$ is given by
\begin{align}
  n_1^{\rm min}=\frac{1}{(m-2)^2}\left\{
{\scriptstyle \sum_j}\sqrt{n_j}-\sqrt{(m-1)\left[
({\scriptstyle \sum_j}\sqrt{n_j})^2-
(m-2){\scriptstyle \sum_j}n_j-
(m-2)
\right]}
\right\}^2
\label{n1_min}
\;,
\end{align}
and it is attained for the following optimal choice of $N_0$
\begin{align}
N_0^{\rm opt}=\frac{1}{(m-1)^2(m-2)^2}\left\{
(m-1){\scriptstyle \sum_j}\sqrt{n_j}-\sqrt{(m-1)\left[
({\scriptstyle \sum_j}\sqrt{n_j})^2-
(m-2){\scriptstyle \sum_j}n_j-
(m-2)
\right]}
\right\}^2-1
\;.
\label{N0_opt}
\end{align}
Substituting \refeq{n1_min} in \refeq{CloFid} one then obtains the
maximum fidelity $F_1^{\rm max}$ achievable for $F_j$ fixed.
Summarizing, if one fixes the fidelities $F_j$ (for $j=2,\dots,m$)
then the thermal photons $n_j$ are given by \refeq{CloFid}, which in
turn individuate the mean photon numbers $N_j$ and $N_1$ of the state
that supports the telecloning via Eqs.~(\ref{noise}), (\ref{n1_min})
and (\ref{N0_opt}). This choice guarantees that the fidelity $F_1$ is
the maximum achievable with the telecloning protocol described above,
thus providing the optimal trade-off between the qualities of the
different clones.
\par
As an example, consider the fully asymmetric $1\rightarrow3$
telecloning and fix the couple of fidelities $F_2$ and $F_3$. Specializing
the formulae above we have that, choosing the state
$|\gr{\Phi}_3\rangle$ such that:
\begin{align}
N_1 =\sqrt{\left(\frac{1}{F_2}-1\right)\left(\frac{1}{F_3}-1\right)}-\frac12 \,, \qquad
N_{2\,(3)} =
\left[
 \sqrt{\frac{1}{F_{3\,(2)}}-1}
-\sqrt{N_1}
\right]^2
\,,
\end{align}
then the fidelity of the first clone is the maximal allowed by our scheme, in formulae:
\begin{align}
  \label{Fid123}
F_1^{\rm max}=\left\{
1+\left[
\sqrt{\frac{1}{F_2}-1}+\sqrt{\frac{1}{F_3}-1}-\sqrt{2\left(
2\sqrt{\left(\frac{1}{F_2}-1\right)\left(\frac{1}{F_3}-1\right)}-1
\right)}
\right]^2
\right\}^{-1}\;.
\end{align}
Notice that $F_1^{\rm max}$ in \refeq{Fid123} is valid iff the fixed
fidelities $F_2$ and $F_3$ satisfy the relation
$F_2\le4(1-F_3)/(4-3F_3)$, which coincides with the optimal relation
given by \refeq{Fid12}. In other words, the optimal bound imposed by
quantum mechanics to $1\rightarrow 2$ telecloning is automatically
incorporated into the bound (\ref{Fid123}) for $1\rightarrow 3$
telecloning of our scheme. When $F_2=F_3=3/5$ (that is, the
bound for an optimal symmetric $1\rightarrow 3$ cloner) we have that
$F_1^{\rm \max}=3/5$, as one may expect from the discussion above
concerning the symmetric cloning case. Remarkably, when $F_2=F_3=2/3$
(that is, the bound for an optimal symmetric $1\rightarrow 2$ cloner)
one has that $F_1^{\rm \max}=1/3>0$. This means that, even if two
optimal clones have been produced, there still remains some quantum
information to produce a non-trivial third clone. A similar situation
occurs for the case of cloning with discrete variables, as pointed out
in Ref.~\cite{IblisdirAsym}.
\par
Similar results occur for the generic $m$ case. In fact, it can be
immediately shown by inspection that substituting $n_j=(m-1)/m$ (that
is, the bound for the noise introduced by an optimal symmetric
$1\rightarrow m$ cloner) in \refeq{n1_min} one obtains   $n_1^{\rm
min}=(m-1)/m$. Hence optimal symmetric cloning is recovered.
Similarly, substituting $n_j=(m-2)/(m-1)$ (that is, the bound for the
noise introduced by an optimal symmetric $1\rightarrow (m-1)$ cloner) in
\refeq{n1_min} one obtains $n_1^{\rm min}=(m-1)/(m-2)$, from which a
fidelity $F_1^{\rm max}=\frac{m-2}{2m-3}>0$ follows. This confirms
that the production of $(m-1)$ optimal clones still leave some quantum
information at disposal to produce an additional non-trivial clone.
An explanation for this effect may be individuated recalling that for large
$m$ the optimal cloner coincides with an optimal measurement on the
original state followed by $m$ reconstruction \cite{cerf}. As a consequence, one
may expect that the production (reconstruction) of $(m-1)$ optimal
clones leaves information ({\em i.e.}, the measurement result) for the
reconstruction of other ones.
\par
A question strictly related to the one faced above, and probably more
significant from an information distribution viewpoint, is the
following.  Suppose that one wants to distribute the information
encoded in the original state by fixing the ratio between the noise
that affects all the $m$ clones, and not by fixing the fidelities of
$(m-1)$ clones. More specifically, suppose that one wants to give the
minimum noise to, say, the first clone ($n_j>n_1$ for every
$j=2,\dots,m$).  Now fix the noise that affects the other clones by
fixing their ratio $q_j$ with respect to the first one, that is
$n_j=q_j\,n_1$.  Then, which is the minimum noise $n_1^{\rm min}$
allowed by our protocol for fixed $q_j$? Solving \refeq{n1_min} for
$n_1$, one may find the following closed expression for $n_1^{\rm min}$ as a
function of $q_j$:
\begin{align}
  n_1^{\rm min}=\frac{m-1}{\left(1+\sum_j\sqrt{q_j}\right)^2-(m-1)\left(1+\sum_jq_j\right)}
\;.
\end{align}
The state $\KPsi$ that provides this optimal result is simply given
by setting the $N_1$ and $N_j$'s obtained by substituting back $n_j^{\rm
  min}=q_j\,n_1^{\rm min}$ in \refeq{N0_opt} and \refeq{noise}.
\par
As a final remark we point out that a general bound for the fidelities
in a fully asymmetric $1\rightarrow m$ cloning of coherent states has
not yet been derived when $m\ge 3$. As a consequence, we cannot judge
if the telecloning process introduced above is in general optimal or
not for $m\ge3$. Nevertheless, there are valuable indications for its
optimality, {\em i.e.} the fact that it is optimal in the case of
$m=2$, and, as we have already pointed out, it is optimal for any $m$
in the symmetric case. In addition, as already mentioned, our
telecloning protocol allows to built a non-trivial additional clone
when $(m-1)$ optimal ones have been produced.
\section{Telecloning in a noisy environment} \label{s:NoisyTLC}
The protocol described in the previous section is referred to the case
of ideal generation and propagation of the states $\KPsi$ as well as
to double-homodyne detection with unit quantum efficiency. In order to
take into account the possible losses and noise in the various steps,
it is useful to reformulate the whole protocol in the phase space.
Consider the characteristic function associated to the states $\KPsi$:
$
\chi[\bmsigma_m](\bmLambda)=\exp\{-\frac12\bmLambda^T\bmsigma_m\bmLambda
\}$. The covariance matrix $\bmsigma_m$ given in \refeq{CovPsi} can be
written accordingly to the following bipartite structure
\begin{align}
\label{BipSigma_m}
\bmsigma_m=
\begin{pmatrix}
  \bmA & \bmC \\
  \bmC^T & \bmB
\end{pmatrix}
\,,
\end{align}
where $\bmA$ is a $2\times2$ matrix corresponding to mode $a_0$, while
$\bmB$ and $\bmC$ are $2m\times 2m$ and $2 \times 2m$ matrices
respectively. Consider now a generic Gaussian POVM, acting on modes
$a_0$ and $b$, defined by a covariance matrix $\bmM$ and a vector of
first moments $\bmX$, {\em i.e.}, $ \chi[\bmM,\bmX](\bmLambda)=\exp\{
-\frac12\bmLambda^T\bmM\bmLambda-i\bmLambda^T\bmX\}$. The case of the
ideal double-homodyne measurement introduced above, corresponds to
\begin{align}
  \bmM=\pp\bmsigma_{\rm in}\pp \,,\qquad \bmX=\pp\overline\bmX +\bmZ \,,
\end{align}
where $\bmsigma_{\rm in}$ and $\overline\bmX$ are the covariance
matrix and the vector of first moments of the input state (mode $b$),
whereas $\bmZ=\{\re z,\im z\}$ is the measurement result [we recall
that $\pp={\rm Diag}(1,-1)$]. Then, the state conditioned to the
result $\bmY$ is given by a Gaussian state with covariance matrix
\begin{align}
  \bmsigma_c=\bmB-\bmC^T(\bmA+\bmM)^{-1}\bmC
\end{align}
and vector of displacements ${\boldsymbol H}=\bmC^T(\bmA+\bmM)^{-1}\bmX$. The
protocol is now completed with the proper generalized local displacement
introduced in the previous section, {\em i.e.}, $U_z =
\bigotimes_{h=1}^m D_h^{\sT}(z)$. Averaging over all the possible outcomes we
finally obtain the following expression for the covariance matrix of
the Gaussian state at the output \cite{besancon}:
\begin{align}
\label{SigmaOut}
\bmsigma=\bmB+\jj^T\pp(\bmA+\bmM)\pp\jj-\jj^T\pp\bmC-\bmC^T\pp\jj
\;,
\end{align}
where $\jj$ is given by the $2\times2m$ matrix $\jj=(\ii,\dots,\ii)$.
\par
As already pointed out in Sec.~\ref{s:NoisySUm1}, if we consider a
realistic scenario for the application of the telecloning protocol, we
must take into account that the generation and the propagation of the
states $\KPsi$ are affected by thermal background and losses. In
particular, concerning propagation we can consider that modes
$a_1,\dots,a_m$ propagate in noisy channels characterized by the same
losses $\Gamma_c$. We may then define an effective propagation time
$\tau_c = \Gamma_c t$ equal for all the clones, while the effective
propagation time $\tau_0 = \Gamma_0 t$ for mode $a_0$ is left
different from $\tau_c$. Consider in fact a scenario in which one has
two distant location (see Fig.~\ref{f:tlc}): the sending station,
where the double-homodyne measurement is performed, and the receiving
station, where the clones are eventually retrieved. The distance
between the two stations can be viewed as a total effective
propagation time $\tau_\tinyT$ which can be written as $\tau_{\tinyT}
=\tau_0+\tau_c$. Then, the choice made above corresponds to the
possibility of choosing at will, for a given $\tau_\tinyT$, which
modes ($a_1,\dots,a_m$ or $a_0$) will be affected by the unavoidable
noise that separates the sending and the receiving station and to
which extent. With a slight abuse of language, we may say that one can
choose whether to put the source of the entangled state $\KPsi$ near
the sending station ($\tau_{\tinyT}=\tau_c$), near the receiving one
($\tau_{\tinyT}=\tau_0$), or somewhere in between. A similar strategy
has been pursued in \cite{welsch} to optimize the CV teleportation
protocol in a noisy environment. In the following, we will see how to
determine both the optimal location and the optimal $\KPsi$ for a
given amount of noise. For the sake of simplicity, the thermal photons
$\mu$ will be taken equal in all the noisy channels. As it is natural to
expect, in the generation process all the modes will be also
considered to be affected by the same amount of noise, characterized
by $\nu$ mean thermal photons. As a consequence, the matrix
$\bmsigma_m$ in \refeq{BipSigma_m} should be substituted by (see,
e.g., Ref.~\cite{Napoli}) its noisy counterpart:
\begin{align}
  \bmsigma_{m,{\rm n}}=\GG^{1/2}\bmsigma_{m,{\rm th}}\GG^{1/2}+(1-\GG)\bmsigma_{\infty,m}
\end{align}
where we have used \refeq{sigma_th}, and defined
\begin{align}
\GG=e^{-\tau_0}\ii\oplus_{j=1}^m\,e^{-\tau_c}\ii
\qquad \bmsigma_{\infty,m}=(\mu+\mbfrac)\ii_{2m}
\end{align}
Performing the calculation explicitly, upon defining
$\gamma_c=e^{-\tau_c}$, $\gamma_0=e^{-\tau_0}$ , $\kappa=\mu+\mbfrac$ and $\zeta=1+2\nu$,
we obtain:
\begin{align}
  \label{sigma_n}
\bmsigma_{m,\rm n}=
\begin{pmatrix}
  \widetilde\bmA & \widetilde\bmC \\
  \widetilde\bmC\,\!^T & \widetilde\bmB
\end{pmatrix}
\,,
\end{align}where $\widetilde\bmA=\zeta\,[\gamma_0\bmcalN_0 + \frac\kappa\zeta(1-\gamma_0)\ii]$, $\widetilde\bmC = \zeta\,\sqrt{\gamma_0\,\gamma_c}\,\bmC$ and
\begin{align}
\widetilde\bmB = \zeta
\left(
\begin{array}{ccccc}
\gamma_c\bmcalN_1 + \frac\kappa\zeta(1-\gamma_c)\ii     & \gamma_c\bmcalB_{1,2} & \ldots          & \gamma_c \bmcalB_{1,m} \\
\gamma_c\bmcalB_{1,2} & \gamma_c\bmcalN_2  + \frac\kappa\zeta(1-\gamma_c)\ii   & \ddots          &  \vdots \\
\vdots        & \ddots        & \ddots          &  \gamma_c\bmcalB_{m-1,m} \\
\gamma_c\bmcalB_{1,m} & \ldots        &\gamma_c \bmcalB_{m-1,m} &  \gamma_c\bmcalN_m + \frac\kappa\zeta(1-\gamma_c)\ii\\
\end{array}
\right)\,.
\end{align}
A non-unit efficiency $\eta$ in the detection stage corresponds to
have the covariance matrix of the double-homodyne detection given by
$\widetilde\bmM=\pp\bmsigma_{\rm in}\pp+\frac12 \Delta\ii$ 
(where $\Delta=\frac{1-\eta}{\eta}$).  Finally, considering an initial
coherent state \cite{note2} and recalling
\refeq{SigmaOut}, we have  
$\widetilde\bmM=\frac12 (1+\Delta)\ii$, whereas  
the covariance matrix of the $m$ output modes now reads:
\begin{align}
  \label{SigmaOutNoise}
  \bmsigma_{\rm
    n}=\widetilde\bmB+\jj^T\pp(\widetilde\bmA+\widetilde\bmM)\pp\jj-\jj^T\pp\widetilde\bmC-\widetilde\bmC\,\!^T\pp\jj
  \;,
\end{align}
which in turn gives the following covariance matrix for the $h$-th clone:
\begin{align}
\label{Reduced_n}
\bmsigma_{h,{\rm n}}=
\left(\frac{1}{F_h}-\frac12\right)
\ii\,.
\end{align}
In the Equation above, $F_h$ represents the fidelity between the $h$-th clone and the
original coherent state:
\begin{align}
\label{FidNoise}
F_h &=\left\{\det\left[\bmsigma_{h,{\rm n}}+\mbfrac\ii
\right]\right\}^{-1/2} \nonumber \\
&= \left\{1+\frac\Delta2+2\kappa+\zeta\left[
\gamma_0\left(N_0+\frac12-\frac\kappa\zeta\right)
+\gamma_c\left(N_h+\frac12-\frac\kappa\zeta\right)
-2\,\sqrt{\gamma_0\,\gamma_c\,N_h(N_0+1)}\right] \right\}^{-1} \,.
\end{align}
\subsection{Optimization of the symmetric protocol} \label{ss:optim}
In order to clarify the implication of the formula (\ref{FidNoise}), let us focus our
attention to the case of symmetric cloning (recall that in this case
$N_1=\ldots=N_m=N$). Upon defining $x=\frac\kappa\zeta-\frac12$,
$\gamma_\tinyT=e^{-\tau_\tinyT}=\gamma_0\gamma_c$ and the following
function
\begin{align}
  \label{f}
\funf=\frac{\gamma_\tinyT}{\gamma_0}(N-x)+\gamma_0(m\,N-x)-2\,\sqrt{\gamma_\tinyT N(m\,N+1)}
\;,
\end{align}
the fidelity reads as follows
\begin{align}
  \label{NFidSym}
F=\left\{\zeta\,\funf+2\,\kappa+1+\frac{\Delta}{2}
\right\}^{-1}\,.
\end{align}
Our aim is now to optimize, for a fixed amount of noise, the
shared state $\KPsi$ and the location of its source between the sending and
the receiving station. Namely, one has to find $N$ and $\gamma_0$
which maximize the fidelity $F$ for $\gamma_\tinyT$, $\kappa$,
$\zeta$, $\Delta$ fixed. This, in turn, means to minimize
$f(N,\gamma_0;x,\gamma_\tinyT)$ for fixed $\gamma_\tinyT$ and $x$.
The domain where to perform the minimization is the region $N>0$ and
$\gamma_T<\gamma_0<1$. We will see that the possibility of varying
$\gamma_0$ will reveal crucial in order to adapt the ideal cloning
protocol, presented in Sec.~\ref{s:tlc}, to a noisy environment.
\par
Calculating the stationary points of $\funf$ one finds:
\begin{align}
  s_1&=\left\{ N=\frac{x}{1-x(m-1)}\,,\,\gamma_0=\sqrt{\frac{\gamma_\tinyT\,x}{x+1}}   \right\}\,, \nonumber \\
 s_2&=\left\{ N=\frac{x}{m[1+x(m-1)]}\,,\,\gamma_0=\sqrt{\frac{\gamma_\tinyT\,(1+m\,x)}{m\,x}}   \right\}\,.
\end{align}
The points $s_1$ and $s_2$ belong to the domain for   $\left\{
\gamma_\tinyT<\frac{x}{x+1}\,, x<\frac{1}{m-1} \right\}$ and for
$\left\{ \gamma_\tinyT<\frac{m\,x}{m\,x+1}\,, \forall x \right\}$
respectively. By evaluating the Hessian matrix associated to
$\funf$, it follows that both $s_1$ and $s_2$ are not extremal
points. As a consequence one has to look for the minimum of
$\funf$ along the boundary of the minimization domain. Three local minima
are found in the three regions parametrized by
$\gamma_0=\gamma_\tinyT$, $\gamma_0=1$ and $N\rightarrow\infty$, whereas the forth extremum is a maximum.
In particular, in the first region the minimum is attained for
\begin{align}
  N=\left\{
  \begin{array}{ll}
   -\dfrac{\gamma_\tinyT}{m\gamma_\tinyT-1} & \;\gamma_T>1/m \\
\dfrac{1}{m(m\gamma_\tinyT-1)} & \;\gamma_T<1/m
  \end{array} \right.
\end{align}
Concerning the second and the third region, one finds that the minima
are located at 
\begin{align}
N=-\frac{\gamma_\tinyT}{m(\gamma_\tinyT-m)}  
\end{align}
and at
\begin{align}
   \gamma_0=\sqrt{\frac{\gamma_\tinyT}{m}}\;,
\end{align} 
respectively. By evaluating
the value of $\funf$ in the minima, one eventually attain the global
maximum $F^{\rm max}$ of the fidelity. A summary of the results is given in
Tab.~\ref{optim}, where we have reintroduced the effective propagation
times $\tau_{\tinyT}$, $\tau_0$ and defined the following quantities:
\begin{align}
F^a &= \frac{2\,m}{
-2-4\,\nu+m\left\{\Delta+2\left[2+\mu+\nu+(\nu-\mu)e^{-\tau_\tinyT}
\right] \right\}} \label{Fa}\,,\\
F^b &= 2\left\{\Delta+2(2+\mu+\nu)-2\,(1+\mu+\nu)e^{-\tau_\tinyT}
\right\}^{-1} \label{Fb}\,,\\
F^c &= \bigg\{
2+\frac\Delta2+2\mu-\sqrt{\frac{e^{-\tau_\tinyT}}{m}}\left[1+\mu+\nu+m(\mu-\nu)
\right]   \bigg\}^{-1} \label{Fc}\,.
\end{align}
\begin{table}[h]
\begin{center}
\begin{tabular}{|c|c|c|c|c|}
\hline $x$ & $\tau_\tinyT$  & $\tau_0^{\rm opt}$ & $N^{\rm opt}$ &
$F^{\rm max}$ \vspace{0cm}\\
\hline\hline
\rule[-4mm]{0mm}{1.1cm} 
$\forall x$ & 
$0<\tau_\tinyT<\ln m$ & 
$\tau_\tinyT$ & 
${\displaystyle \frac{1}{m(m\,e^{-\tau_\tinyT}-1)}}$ & 
$F^a$ \\
\hline
\rule[-4mm]{0mm}{1.1cm}
$-\frac12<x<0$ &
$\tau_\tinyT>\ln m$ &
$\mbfrac(\tau_\tinyT+\ln m)$ & 
$N\rightarrow\infty$ &
$F^c$ \\
\hline
\multirow{2}*{\rule[-2mm]{0mm}{1cm} $0<x<\frac{1}{m-1}$} 
\rule[-4mm]{0mm}{1.1cm}& 
${\displaystyle \ln m<\tau_\tinyT<\ln\left[\frac{(1+x)^2}{m\,x^2}\right]}$ & 
$\mbfrac(\tau_\tinyT+\ln m)$ & 
$N\rightarrow\infty$ &
$F^c$ \\
\cline{2-5} & \rule[-4mm]{0mm}{1.1cm}
$\tau_\tinyT>\ln \frac{(1+x)^2}{m\,x^2}$ & 
$\tau_\tinyT$ & 
${\displaystyle \frac{e^{-\tau_\tinyT}}{1-m\,e^{-\tau_\tinyT}}
}$ &
$F^b$ \\
\hline
\rule[-4mm]{0mm}{1.1cm}
$x>\frac{1}{m-1}$ & 
$\tau_\tinyT>\ln m$ & 
$\tau_\tinyT$ & 
${\displaystyle \frac{e^{-\tau_\tinyT}}{1-m\,e^{-\tau_\tinyT}}}$ &
$F^b$ \\
\hline
\end{tabular}
\end{center}
\caption{ Values of the optimized $N^{\rm opt}$ and $\tau_0^{\rm
opt}$ for fixed values of $\tau_\tinyT$ and $x$. The value reached
by the fidelity $F^{\rm max}$ for these optimal choices is given
in the last column. \label{optim}}
\end{table}
\par
\begin{figure}[h]
\vspace{2cm}
\setlength{\unitlength}{0.4cm} 
\centerline{%
\begin{picture}(15,0)
\put(0,0){\makebox(-3,0)[c]{\epsfxsize=5cm\epsffile{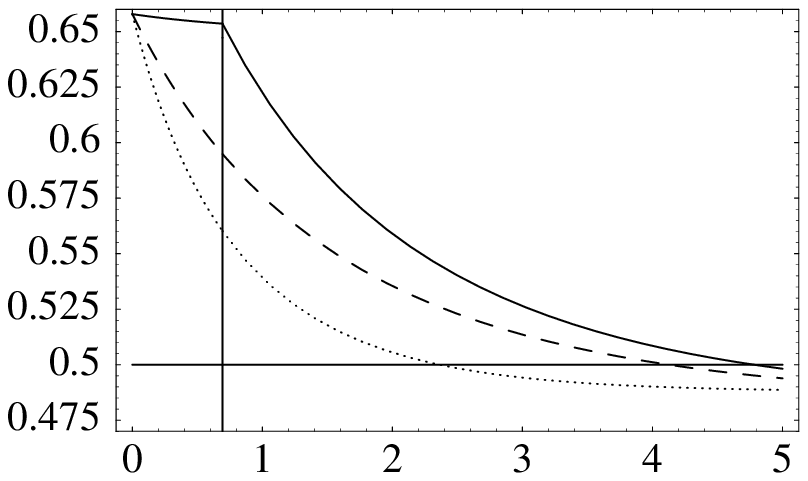}}}
\put(3,-4.4){$\tau_\tinyT$}
\put(-8.5,3.8){$F^{\rm max}$}
\put(2,2.8){\footnotesize $m=2$}
\end{picture}
\begin{picture}(15,0)
\put(0,0){\makebox(-3,0)[c]{\epsfxsize=5cm\epsffile{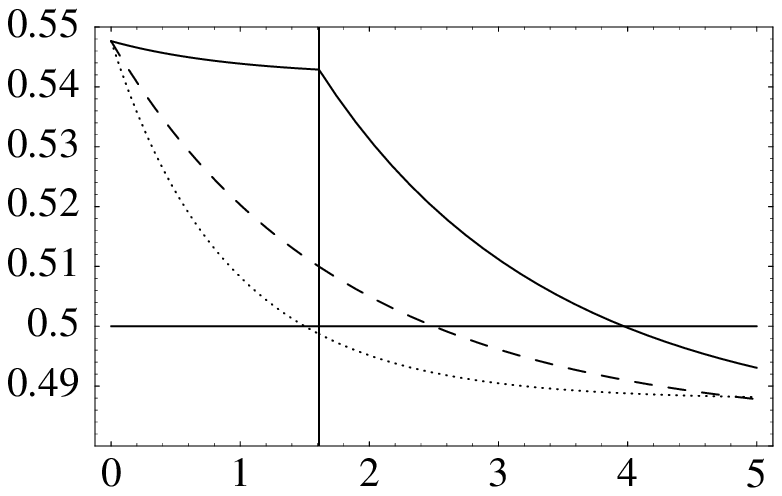}}}
\put(3,-4.4){$\tau_\tinyT$}
\put(-8.5,3.8){$F^{\rm max}$}
\put(1.6,2.5){\footnotesize $m=5$}
\end{picture}
\begin{picture}(-4,0)
\put(0,0){\makebox(-3,0)[c]{\epsfxsize=5cm\epsffile{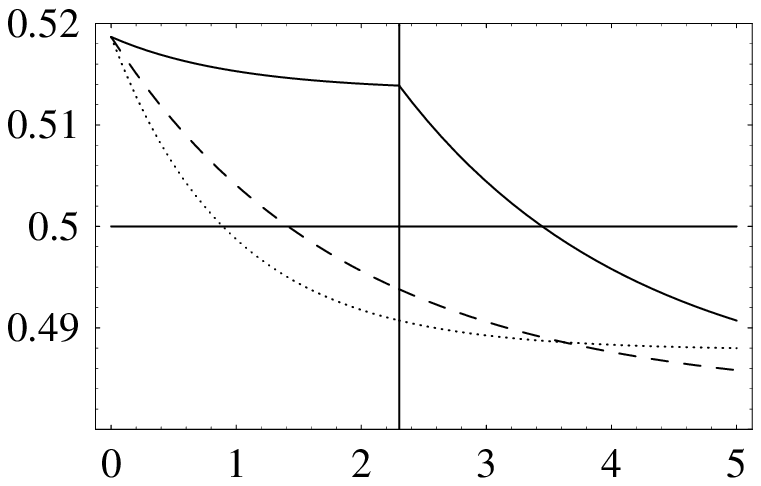}}}
\put(3,-4.4){$\tau_\tinyT$}
\put(-8.5,3.8){$F^{\rm max}$}
\put(1.3,2.5){\footnotesize $m=10$}
\end{picture}
}
\vspace{1.8cm}
\caption{Comparison of the fidelities given in Eqs.~(\ref{FeasF1}) (dotted line) and (\ref{FeasF2}) (dashed line) with the one
  given in Tab.~\ref{optim} (solid line). As an example, we have
  chosen the following parameters: $\mu=0.03$, $\nu=0.01$ and
  $\Delta=0.02$ ($x=1/51$).  The plots are referred to the case
  $m=2,5,10$ and the vertical line corresponds to $\tau_\tinyT=\ln 2$,
  $\tau_\tinyT=\ln 5$ and $\tau_\tinyT=\ln 10$ respectively.
  Accordingly to Tab.~\ref{optim}, the optimal fidelity is given by
  \refeq{Fa} and \refeq{Fc} at the left and at the right of the
  vertical lines, respectively.}
\label{f:OptVsFeas}
\end{figure}
\begin{figure}[h]
\vspace{1.5cm}
\setlength{\unitlength}{0.4cm} 
\centerline{%
\begin{picture}(15,0)
\put(0,0){\makebox(-3,0)[c]{\epsfxsize=5cm\epsffile{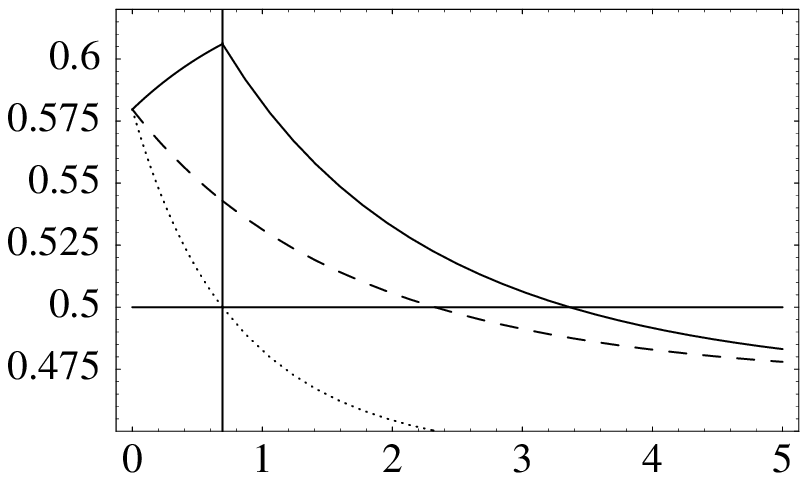}}}
\put(3,-4.4){$\tau_\tinyT$}
\put(-8.5,3.8){$F^{\rm max}$}
\put(2,2.8){\footnotesize $m=2$}
\end{picture}
\begin{picture}(15,0)
\put(0,0){\makebox(-3,0)[c]{\epsfxsize=5cm\epsffile{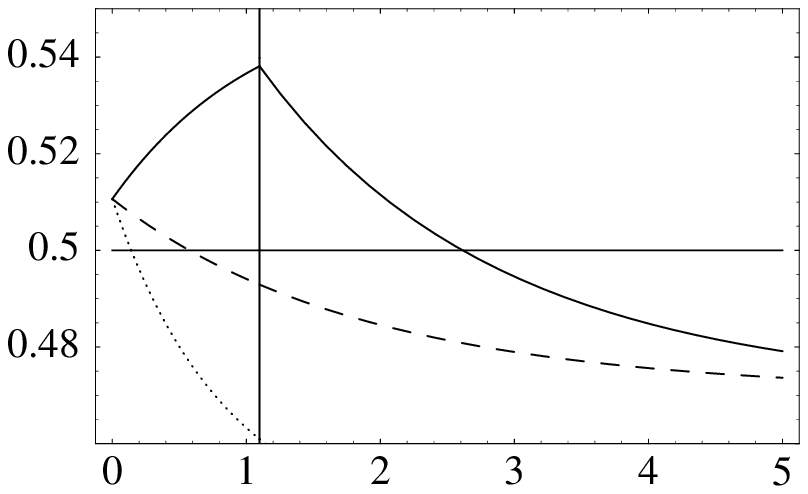}}}
\put(3,-4.4){$\tau_\tinyT$}
\put(-8.5,3.8){$F^{\rm max}$}
\put(1.6,2.8){\footnotesize $m=3$}
\end{picture}
\begin{picture}(-4,0)
\put(0,0){\makebox(-3,0)[c]{\epsfxsize=5cm\epsffile{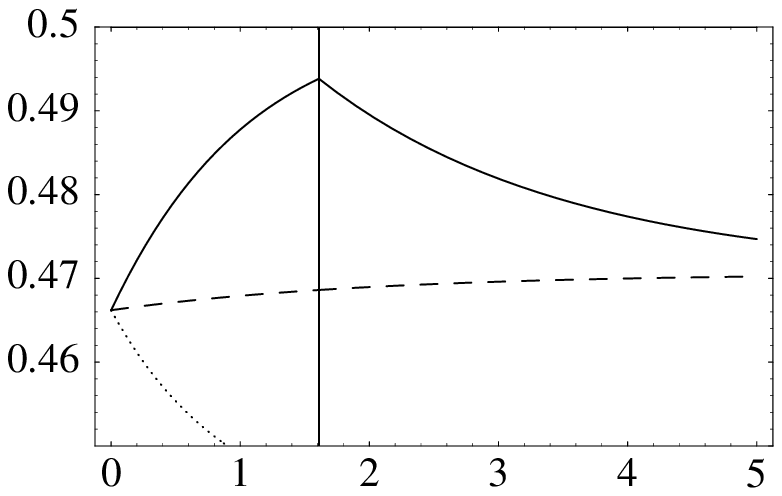}}}
\put(3,-4.4){$\tau_\tinyT$}
\put(-8.5,3.8){$F^{\rm max}$}
\put(1.3,2.5){\footnotesize $m=5$}
\end{picture}
}
\vspace{1.8cm}
\caption{Comparison of the fidelities given in Eqs.~(\ref{FeasF1}) (dotted line) and (\ref{FeasF2}) (dashed line) with the one
  given in Tab.~\ref{optim} (solid line). As an example of the case
  $\mu<\nu$, we have chosen the following parameters: $\mu=0.05$,
  $\nu=0.2$ and $\Delta=0.05$ ($x=-3/28$).  The plots are referred to
  the case $m=2,3,5$ and the vertical line corresponds to
  $\tau_\tinyT=\ln 2$, $\tau_\tinyT=\ln 3$ and $\tau_\tinyT=\ln 5$
  respectively.  Accordingly to Tab.~\ref{optim}, the optimal fidelity
  is given by \refeq{Fa} and \refeq{Fc} at the left and at the right
  of the vertical lines, respectively.}
\label{f:OptVsFeas2}
\end{figure}
\begin{figure}[h]
\vspace{1.5cm}
\setlength{\unitlength}{0.4cm} 
\centerline{%
\begin{picture}(15,0)
\put(0,0){\makebox(-3,0)[c]{\epsfxsize=5cm\epsffile{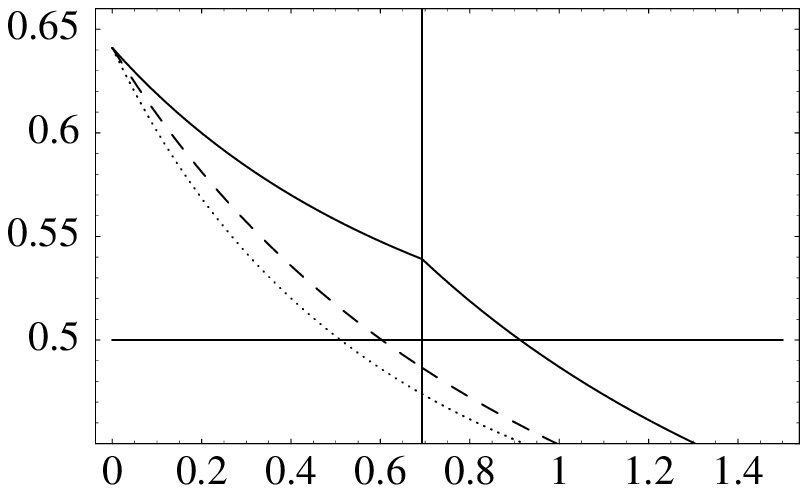}}}
\put(3,-4.4){$\tau_\tinyT$}
\put(-8.5,3.8){$F^{\rm max}$}
\put(2,2.8){\footnotesize $m=2$}
\end{picture}
\begin{picture}(-5,0)
\put(0,0){\makebox(-3,0)[c]{\epsfxsize=5cm\epsffile{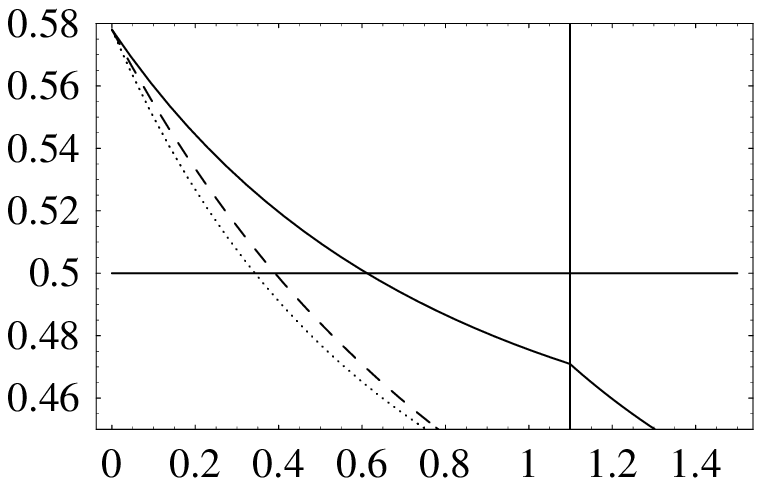}}}
\put(3,-4.4){$\tau_\tinyT$}
\put(-8.5,3.8){$F^{\rm max}$}
\put(1.7,2.6){\footnotesize $m=3$}
\end{picture}
}
\vspace{1.8cm}
\caption{Comparison of the fidelities given in Eqs.~(\ref{FeasF1}) (dotted line) and (\ref{FeasF2}) (dashed line) with the one
  given in Tab.~\ref{optim} (solid line). As an example, we have
  chosen the following parameters: $\mu=0.6$, $\nu=0.01$ and
  $\Delta=0.1$ ($x=59/102$).  The plots are referred to the case
  $m=2,3$ and the vertical line corresponds to $\tau_\tinyT=\ln 2$
  and $\tau_\tinyT=\ln 3$ respectively.  Accordingly to
  Tab.~\ref{optim}, for $m=2$ the optimal fidelity is given
  by \refeq{Fa} and \refeq{Fc} at the left and at the right of the
  vertical line, respectively. For $m=3$, it is instead
  given by \refeq{Fb} at the right of the vertical line. }
\label{f:OptVsFeas3}
\end{figure}

An inspection of Tab.~\ref{optim} shows very interesting features of
the telecloning protocol in presence of noise. It is immediate to
recognize that the optimal value $N^{\rm opt}$ is significantly
different from the optimal value in the ideal case [\refeq{NOptId}].
As a matter of fact $N^{\rm opt}$ is divergent in some cases.
Remarkably, in the optimization of $N$ and $\tau_0$ the homodyne
detection efficiency $\Delta$ plays no role, whereas the thermal
noises $\mu$ and $\nu$ introduce a dependence on $x$, {\em i.e.} only
on their ratio. Furthermore, one may note that what we have called the
best location of the source (that is $\tau_0^{\rm opt}$) is never
given by the simple choices $\tau_0=0$ or $\tau_0=\tau_\tinyT/2$. In
order to clarify this point let us first consider the case $\tau_0=0$,
which can be physically implemented by homodyning mode $a_0$
immediately after the generation of $\KPsi$, and then letting the
other modes propagate to the receiving station where they are
eventually displaced. An immediate calculation shows that in this case
the fidelity (\ref{NFidSym}) is maximized for $N^{\rm
  opt}=1/m(m\,e^{\tau_\tinyT}-1)$ and is given by
\begin{align}
  F^{\rm max}(\tau_0=0)=\frac{2\,m}{2\,e^{\tau_\tinyT}\left[1+m(\mu-\nu)+2\,\nu \right]-m\left[\Delta+2(2+\mu+\nu)\right]}\;.
\label{FeasF1}
\end{align}
Concerning the case $\tau_0=\tau_\tinyT/2$, whose physical
implementation simply means to put the source of $\KPsi$ in the middle
of the transmission line, one has that the fidelity is maximized for
$N^{\rm opt}=1/m(m-1)$ and reads
\begin{align}
F^{\rm max}(\tau_0=\tau_\tinyT/2)
=\frac{2\,m}{m(4+\Delta+4\,\mu)-2e^{-\tau_\tinyT/2}\left[1+2\,\nu+2\,m(\mu-\nu)\right]}\;.
\label{FeasF2}
\end{align}
Notice that only in this case the optimization over $N$ leads to the
same $\KPsi$ as in the ideal case (see \refeq{NOptId}). A comparison
of the last two instances with the optimal one, shows how
significantly the choice of $\tau_0$ affects the value of the clones'
fidelity.  In Figs.~\ref{f:OptVsFeas}, \ref{f:OptVsFeas2} and
\ref{f:OptVsFeas3} we compared the two fidelities given in
Eqs.~(\ref{FeasF1}) and (\ref{FeasF2}) with the one given in
Tab.~\ref{optim} (see captions for details). We clearly see that the
optimized fidelity is much larger then the other two, thus providing a
cloning beyond the classical limit for higher propagation times
$\tau_\tinyT$. As it is apparent from Fig. \ref{f:OptVsFeas3} we have 
$F^b < \frac12$ $\forall \tau_\tinyT$. Indeed, it can be shown
analitically that $F^b<\frac12$ in any regime for which $F^{\rm
max}=F^b$. 
\par
Besides what we pointed out above, the most striking feature of the
proposed telecloning protocol is that it saturates the bound for
optimal cloning even in the presence of losses, for propagation times
$\tau_\tinyT<\ln m$, hence divergent as the number of modes increases.
More specifically, consider the first row in Tab.~\ref{optim} and set
$\mu=\nu=\Delta=0$. Then, one has that for $\tau_\tinyT<\ln m$ the
maximum fidelity is given by $F^{\rm max}=m/(2m-1)$. That is, the
optimal fidelity for a symmetric cloning can still be attained,
carefully choosing $N$ and $\tau_0$. Such a result cannot be achieved
letting the input state propagate directly to the receiving station
and then cloning it locally \cite{besancon}. Thus, in the context of our protocol the
entangled resource significantly enhances the capacity of distributing
quantum information. This is due to the fact that the transmission
through a lossy channel of an unknown coherent state irreversibly
degrades the information encoded in it, thus avoiding the local
construction of optimal clones at the receiving station. On the other
hand, multimode entanglement is robust against this type of noise and,
even if decreased along the transmission line, it is still sufficient
to provide optimal cloning. Actually, there is no need of an infinite
amount of entanglement to perform an optimal telecloning process.
\par
As concern the case of higher transmission times, {\em i.e.}
$\tau_\tinyT>\ln m$, the fidelity reads (again for $\mu=\nu=\Delta=0$)
\begin{align}
  F^{\rm max}=\left(2-\sqrt{\frac{e^{-\tau_\tinyT}}{m}}\right)^{-1}.
\label{LossyFLong}
\end{align}
Eq.~(\ref{LossyFLong}) shows that the fidelity is always greater then the
classical bound $F=\frac12$, which in turn means that the state used
to support the protocol is entangled for any $\tau_\tinyT$.  This is
reminiscent of the result already pointed out in
Sec.~\ref{s:NoisySUm1}, where the full inseparability has been proved
for any $\tau_\tinyT$ for $m=2$ (notice that $\tau_0=\tau_\tinyT/2$ in case
of Sec.~\ref{s:NoisySUm1}). Here, we proved that the same conclusion
is valid for any $m$.
\par
Another interesting feature in the case $\tau_\tinyT<\ln m$ is that
$F^{\rm max}$ does not depend on $\tau_\tinyT$ if $\mu=\nu$. Indeed,
it turns out that for $\mu=0$ and $\nu\neq0$ it is better to let the
entangled resource propagate (up to $\tau_T=\ln m$) instead of using
it immediately after the generation. This effect may be naively
understood by considering that the entangled state generated for
$\nu\neq0$ is mixed and, as consequence, the propagation in a purely
dissipative environment acts like a sort of purification process on
it. As it is apparent from \refeq{Fa} this effect is present whenever
$\mu<\nu$ (see also Fig.~\ref{f:OptVsFeas2}).
\par
Finally, a comment is needed concerning the scaling of the fidelity with respect
to the number of modes $m$. We have already pointed out that for the
case $\mu=\nu=\Delta=0$ the fidelity remains optimal for times
$\tau_\tinyT$ diverging with the number of modes. However, when
thermal noise is added ($\mu,\nu,\Delta\neq0$) the fidelity goes below
the classical value $F=\frac12$ for times $\tau_\tinyT$ that become
smaller as $m$ increases, as it is apparent from
Figs.~\ref{f:OptVsFeas}, \ref{f:OptVsFeas2} and \ref{f:OptVsFeas3}.
Indeed, this is consistent with the fact that the optimal fidelity itself
approaches the classical value $F=\frac12$ as $m$ increases. Hence, even a
small amount of thermal noise is enough to cancel the benefits
due to quantum entanglement.
\section{Conclusions}\label{esco}
In this paper we have dealt with the properties and applications of a
class of multimode states of radiation, the coherent states of group
$\rmSU (m,1)$, which represent a potential resource for multiparty
quantum communication, as recent theoretical and experimental
investigation have shown. In particular, the common structure of these
multimode states allowed to consider a $1\rightarrow m$ telecloning
scheme in which a generic coherent state of $\rmSU (m,1)$ plays the
role of entangled resource. Exploiting the possible asymmetry of
$\rmSU (m,1)$ coherent states we have suggested the first example, in
the framework of CV systems, of a fully asymmetric $1\rightarrow m$
cloning and have found the optimal relation, within our scheme,
between the different fidelities of the clones. In particular, we have
shown that when $(m-1)$ optimal clones are produced (accordingly to
the general bound imposed by quantum mechanics), there still remains
some quantum information at disposal.  In fact, our protocol is able
to use the remaining information to realize a non-trivial $m$-th
clone. Our asymmetric scheme is aimed at the distribution of quantum
information among many parties \cite{qid}, and may find application
for quantum cryptographic purposes \cite{gisin}.
\par
In view of possible applications of our protocol in realistic
situations, we have considered the effects of noise in the various
stages of the protocol, {\em i.e.} the presence of thermal photons in
the generation process, thermal noise and losses during propagation,
and non-unit efficiency in the detection. We have derived the
fidelities of the clones as a function of the noise parameters, which
in turn allowed for adaptive modification of the protocol to face the
detrimental effects of noise. In particular, we have shown that the
optimal entangled resource in the presence of noise is significantly
different from the one in the ideal case. Also the location of the
source plays a prominent role. In fact, we have demonstrated that the
optimal location is neither in the middle between the sender and the
receiver, nor at the sender station. A striking feature of the
optimized protocol is that, even in the presence of losses along the
propagation line, the clones' fidelity remains maximal, a result which
is not achievable by means of direct transmission followed by local
cloning. This happens for propagation times that diverge as the number
of modes increases. We then conclude that our optimized telecloning
protocol is robust against noise.
\section*{Acknowledgments}
The authors are grateful to S.~Olivares for fruitful discussions. This work has
been partially supported by MIUR (FIRB RBAU014CLC-002) and by INFM
(PRA-CLON).



\begin{thebibliography}{999}
\bibitem{note1} Notice that state (\ref{shared}) does not belong to
  the class of the coherent states of $\rmSU(m,1)$ anymore.
  Nevertheless being the additional squeezing operations local, then
  all the entanglement used to support the telecloning process is
  furnished by $\KPsi$.
\bibitem{note2} The protocol can be straightforwardly
  generalized to squeezed states as in the ideal case.
\end{thebibliography}
\end{document}